\def\year{2022}\relax
\documentclass[letterpaper]{article} % DO NOT CHANGE THIS
\pdfoutput=1
\usepackage{aaai22}  % DO NOT CHANGE THIS
\usepackage{times}  % DO NOT CHANGE THIS
\usepackage{helvet}  % DO NOT CHANGE THIS
\usepackage{courier}  % DO NOT CHANGE THIS
\usepackage[hyphens]{url}  % DO NOT CHANGE THIS
\usepackage{graphicx} % DO NOT CHANGE THIS
\usepackage{natbib}  % DO NOT CHANGE THIS AND DO NOT ADD ANY OPTIONS TO IT
\usepackage{caption} % DO NOT CHANGE THIS AND DO NOT ADD ANY OPTIONS TO IT
\DeclareCaptionStyle{ruled}{labelfont=normalfont,labelsep=colon,strut=off} % DO NOT CHANGE THIS
\usepackage{algorithm}
\usepackage{algorithmic}
\usepackage{amsmath}
\usepackage{amsfonts}
\usepackage{newfloat}
\usepackage{listings}
\floatstyle{ruled}
\newfloat{listing}{tb}{lst}{}
\floatname{listing}{Listing}
\usepackage{upgreek}

\title{Adaptive Energy Management for Self-Sustainable Wearables in Mobile Health}
\author{
    Dina Hussein, Ganapati Bhat, Janardhan Rao Doppa
}
\affiliations{
    School of EECS, Washington State University, Pullman, WA 99164\\
    \{dina.hussein, ganapati.bhat, jana.doppa\}@wsu.edu
}

\begin{document}

\maketitle

\begin{abstract}

Wearable devices that integrate multiple sensors, processors, and communication technologies have the potential to transform mobile health for remote monitoring of health parameters. However, the small form factor of the wearable devices limits the battery size and operating lifetime. As a result, the devices require frequent recharging, which has limited their widespread adoption. Energy harvesting has emerged as an effective method towards sustainable operation of wearable devices. %by harnessing the energy available in the environment. 
Unfortunately, energy harvesting alone is not sufficient to fulfill the energy requirements of wearable devices. This paper studies the novel problem of {\em adaptive energy management} towards the goal of self-sustainable wearables by using harvested energy to supplement the battery energy and to reduce manual recharging by users. %One of the key challenges is to ensure that the manual recharging does not interfere with critical periods of user activity (e.g. gait monitoring in a patient with Parkinson's disease). 
To solve this problem, we propose a principled algorithm referred as {\em AdaEM}. There are two key ideas behind AdaEM. First, it uses machine learning (ML) methods to learn predictive models of user activity and energy usage patterns. These models allow us to estimate the potential of energy harvesting in a day as a function of the user activities. Second, it reasons about the uncertainty in predictions and estimations from the ML models to optimize the energy management decisions using a dynamic robust optimization (DyRO) formulation. We propose a light-weight solution for DyRO to meet the practical needs of deployment. %about managing the harvested energy so that the interval between charging cycles is maximized. 
We validate the AdaEM approach on a wearable device prototype consisting of solar and motion energy harvesting using real-world data of user activities. Experiments show that AdaEM achieves solutions that are within 5\% of the optimal with less than 0.005\% execution time and energy overhead. 
\end{abstract}

\section{Introduction}
% Motivate that healthcare costs are increasing. Wearables can help

The increasing cost of healthcare ~\cite{dieleman2020us} and its growing need due to chronic diseases and aging~\cite{dorsey2016moving} motivates the need for remote health-monitoring~\cite{espay2016technology}. Towards this goal, wearable devices that integrate multiple physiological sensors, processors, and communication technologies to monitor the user's health have emerged as a promising technology~\cite{daneault2018could}. For example, in patients diagnosed with cardiac arrhythmia, we can use the measurements from sensors to predict the likelihood of a cardiac failure. The design of wearable devices is optimized for user comfort using emerging technologies including flexible hybrid electronics \cite{khan2016flexible}, as they need to be worn for an extended duration. 
As a result, wearable devices are typically light-weight with a small form-factor which severely limits the size of the available battery. The small battery, in turn, limits the operating life, leading to the need for frequent recharging, reduced usability, and adoption ~\cite{johansson2018wearable}. 
Therefore, there is a strong need for approaches that enable sustainable operation of wearable devices and minimize recharging requirements.

% Energy harvesting is useful here but it is highly stochastic. Furthermore, it is not sufficient to power the devices by itself. 
Energy harvesting (EH) \cite{kansal2007power,vigorito2007adaptive,9519688} from ambient sources has emerged as a promising solution to enable self-sustainable wearable devices. We can integrate various EH modalities including photovoltaic~(PV) cells, piezoelectric sensors, and thermoelectric generators to harvest energy from ambient light, user motion, and body heat, respectively.
% Recent studies have shown that wearable devices can harvest up to XX Joules of energy per hour using these modalities \cite{}.
The harvested energy from ambient sources can be used to supplement the battery of the device and prolong the interval between manual recharges. The overall goal of energy management (EM) is to maximize the utilization of harvested energy and to reduce the frequency of manual recharging. EM decisions include the timing and duration of manual recharging. Decision-making for EM is challenging for the following reasons: {\bf 1)} energy availability from ambient sources is highly stochastic. EM decision-making should account for this inherent uncertainty so that the battery is not exhausted completely; {\bf 2)} the wearable device must not miss any critical events, such as falls, during recharging periods; and {\bf 3)} recharging during intervals of abundant energy availability can lead to poor utilization of harvested energy. Therefore, EM algorithms must learn the usage patterns of the users and the EH potential to optimally schedule manual recharging.

% Proposed approach in this paper

This paper proposes a novel and principled AI-based approach referred to as {\em AdaEM} to solve the above-mentioned challenges of energy management. AdaEM relies on two key ideas. First, we use machine learning (ML) techniques to predict user activities, energy usage patterns, and EH availability in the future. These predictions are fed as input for EM decision-making. Second, to reason about the worst-case uncertainty in ML predictions and ambient energy harvest, we formulate a {\em dynamic robust optimization (DyRO)} problem to make EM decisions. Due to limited compute and energy resources in wearables, we cannot use expensive methods to repeatedly solve DyRO problem instances for decision-making. Therefore, we propose a highly effective light-weight approach to solve DyRO instances for real-world deployment. We validate our AdaEM approach on a real wearable device prototype consisting of light and motion EH. We start by characterizing the potential of energy harvest on the prototype.
The characterization data is then combined with the user activity data from publicly available datasets to determine the available energy in day-to-day activities of the users. Finally, we apply AdaEM on the user data to maximize the interval between recharging operations without missing critical activities.
Our results show that AdaEM achieves
solutions that are within 5\% of the optimal solution with less than 0.005\% execution time and energy overhead on real wearables.
Moreover, comparison with a realistic baseline shows that AdaEM has significantly lower energy constraint violations while providing the required level of accuracy for the target application.
% XX\% accuracy with less than 1\% runtime and energy overhead on real wearables. %on this real wearable prototype.

\vspace{0.5ex}
\noindent {\bf Contributions:} The key contribution of this paper is the development and evaluation of the AdaEM algorithm for energy management in wearable devices for mobile health.
\begin{itemize}
\setlength\itemsep{0em}

\item Specifying the novel AI-based energy management problem for self-sustainable wearable devices via energy harvesting, which has the potential for high social impact.

\item Development of a principled dynamic robust optimization formulation for decision-making to maximize the utilization of the harvested energy. %DyRO reasons about the uncertainty in ML predictions. 
\item A light-weight and highly effective algorithm for repeatedly solving DyRO instances for decision-making to meet the needs of real-world deployment.

%Novel algorithm to schedule recharging operation of wearable devices without missing critical activities and maximizing the utilization of the harvested energy

\item Experimental evaluation on a real wearable prototype with data from five users over five years to show the benefits of the AdaEM algorithm in achieving sustainable operation with negligible overhead. 
We also release the source code on GitHub at: {\ttfamily https://github.com/gmbhat/adaEM}
\end{itemize}

\section{Related Work}
% Motivation for wearables and their applications
Research and development of wearable devices has increased in recent years due to their applications in health and activity monitoring \cite{livmor,dempsey2015teardown}.
% Examples of wearable devices range from heart monitoring devices like the LIVMOR Halo~\cite{livmor} to fitness trackers such as Apple Watch~\cite{dempsey2015teardown}. 
% These devices monitor vital signs of the user and provide feedback on their health status or in improving their activity levels.
% The availability of this information has led to adoption of wearable devices by fitness enthusiasts
However, their adoption by the medical community is limited. 
One of the primary reasons for the limited adoption is the frequent recharging requirements of most, if not all, wearable devices~\cite{ozanne2018wearables}. 
Therefore, recent research in wearable devices has focused on addressing the energy limitations of wearable devices~\cite{chong2019energy}.

% Related work on energy harvesting modalities
Energy harvesting and management has emerged as the most promising technique to overcome the energy constraints in wearable devices.
% Specifically, EH methods harness the energy available in the environment by using the appropriate transducers in the wearable device.
Common sources of energy include ambient light, user motion, and body heat. Out of these sources, ambient light has the highest power density and efficiency. For instance, prior studies have shown that ambient light can provide up to 
0.1 mW/cm$^2$ and 100 mW/cm$^2$ in outdoor and indoor conditions, respectively~\cite{valenzuela2008energy}.
Motion EH has received increased attention recently due to its applicability in activity monitoring devices~\cite{mitcheson2008energy}. 
% In particular,  mounted on the joints harvest energy whenever the user moves their limbs. 

% Radio frequency energy harvesting captures the energy in ambient electromagnetic waves, however, it suffers from low efficiency and energy densities. To overcome this, some approaches use dedicated transmitters to enable wireless power transfer~\cite{}. While this approach is useful, it is limited to indoor environments and short distances due to the attenuation in wireless signals with distance.

% Energy neutral operation work. Kansal, vigorito, preact etc
Harvesting energy from ambient sources necessities development of algorithms that manage the harvested energy effectively. This is critical because the ambient energy is not available at all times during the day. As such, the device must have sufficient energy reserves to operate effectively when harvested energy is not available.
To this end,~\citet{kansal2007power} proposed the concept of energy neutral operation using linear programming, whereby the energy used in a given period (e.g. a day) is equal to the energy harvested in that period. 
% To achieve this, the authors model the available energy for the device and formulate a linear programming problem to calculate the active time of the device throughout the day.
Recent work has also used dynamic programming and control-theoretic methods to achieve energy neutral operation~\cite{vigorito2007adaptive,bhat2017near,geissdoerfer2019getting,yamin2021diet}.

% Charging optimizations and work in the AI community
ML methods have also been used for energy harvest prediction and allocation to achieve energy neutral operation. For instance, neural networks and other predictors have been used to estimate light energy available in the future~\cite{barrera2020solar,dhillon2020solar,shresthamali2017adaptive}. The energy estimates are then used to decide the energy allocation and duty cycle in the future.
%Similarly, reinforcement learning has been used to learn the energy harvesting patterns and make duty cycle decisions for the wearable device~\cite{shresthamali2017adaptive}.
% Use of these ML and AI methods help the wearable device in adapting to changes in the environmental conditions while maintaining energy neutral operation.
These prior methods don't address our general adaptive EM problem. In contrast, we formulate a \textit{dynamic robust optimization} to make EM decisions by accounting for uncertainty in the harvested energy.
We propose a light-weight solution, AdaEM, to make EM decisions at runtime. %Our results show that AdaEM makes decisions that are within 5\% of the optimal Oracle with less than 1\% overhead.

% Application of the DyRO on user data shows that 
% For instance, if the harvested energy on a given day is close to zero, energy neutral operation will lead to the device being turned off for majority of the day.
% This can have potentially catastrophic consequences if the device shuts down when the user has a life threatening event. 
%  Therefore, there is a need for approaches that make use of harvested energy and intelligent manual recharging to enable sustainable operation.
% Operation on harvested energy alone is not possible at the moment, so we need to optimize for the charging cycles
\begin{figure*}[t] 
    \centering
    \includegraphics[width=0.82\linewidth]{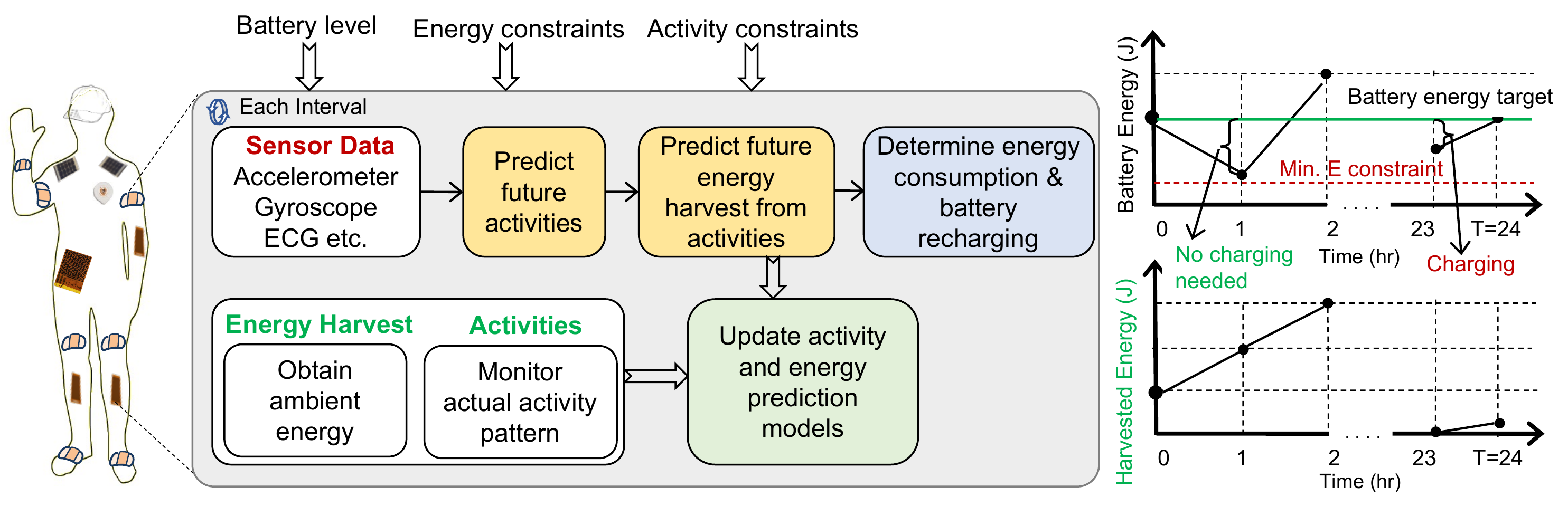}
    \caption{Overview of the system for sustainable wearable health monitoring via adaptive energy management.}
    \label{fig:overview}
\end{figure*}
% Proposed formulation
\section{Background and Preliminaries}
This section first provides a brief background on emerging technologies in form-factor and EH for wearable devices, with detailed explanations in the Appendix.
Then, we describe challenges and opportunities for the widespread adoption of wearables using these technologies.

\noindent\textbf{Flexible hybrid electronics:}
% Early implementations wearable devices included rigid components on a rigid printed circuit board. However, this form factor is not suitable for long term usage because it does not conform to the shape of the body, making it uncomfortable. 
% Indeed, recent user studies have shown that comfort is one of the most common reasons why users stop using rigid wearable devices~\cite{ozanne2018wearables}.
Flexible electronics can enable comfortable wearables through the use of materials that are fully bendable and rollable.
However, the performance capabilities of fully flexible electronics are low compared to conventional CMOS devices.
To address these limitations, recent research has proposed flexible hybrid electronics~(FHE)~\cite{khan2016flexible}.
FHE uses rigid components on a flexible substrate to implement a device that conforms to the shape of the user's body. By using rigid components for processing and sensing, FHE devices are able to utilize the performance advantages provided by CMOS technology and form factor advantages of the substrate.

% \vspace{1mm}
\noindent\textbf{Light EH:} 
% Photo-voltaic cells have been used to harvest energy for a number of years. 
% Recently, s
Small form-factor flexible PV cells have shown potential for use in wearable devices~\cite{park2009flexible}.
These PV cells can be easily integrated into wearable devices including fabrics, hats, and jackets. Therefore, they are an important step towards sustainable operation and adoption of wearable devices.

\noindent\textbf{Motion EH:} Motion EH is another promising technology for wearable devices as motion energy harvesters can be easily integrated with human activities~\cite{mitcheson2008energy}. For example, piezoelectric harvesters can be used on knees and elbows to harvest energy whenever the user moves. 
Some studies have also used motion EH as a feature in performing activity classification~\cite{khalifa2017harke}.
As a result, motion EH can act as a sensor and an energy source. 
% Moreover, motion energy harvesting can provide on-demand energy for wearable applications by harvesting energy whenever the user moves and there is a need for classifying their activities.

% and other advancements such as low power communication and sensing

\subsection{Challenges \& Opportunities for Wearables}
Despite the development of the above technologies, the widespread adoption of wearable devices has been limited to laboratory studies and smartwatches.
One of the primary reasons for the lack of adoption of wearable devices is frequent recharging requirements along with lack of comfort of rigid devices. 
% Recent research on addressing these challenges typically focus on a single challenge at a time. For instance, EH approaches typically do not consider FHE as part of the energy harvesting or management.
% Similarly, research on FHE and wearable health applications do not take into account energy harvesting and management. This results in various research silos with limited cross-application of methods.
To address these challenges, we take a holistic view of the problem.
Specifically, our problem formulation considers the application requirements, user activity patterns, and EH together to enable optimal operation of wearable devices.
The formulation also includes unique properties of FHE, such as bending, so that the changes in EH are accounted correctly~\cite{park2017pv}.

Figure~\ref{fig:overview} shows an overview of the system for sustainable wearable health monitoring.
The left side shows an illustration of a user wearing number of sensors and EH modules. The data obtained from the sensors is used to identify the user's activities and health status.
During each interval of operation, the system takes the current battery level, energy constraints, and activity constraints as the input.
The energy constraints may specify that the battery must be maintained above a certain minimum level to ensure sufficient reserves in case of an emergency. Similarly, the activity constraints specify a set of activities that have the highest priority for the user.
The first step in the EM is to predict future activities using the sensor data. The activities are then used to predict future energy requirements and harvest. Finally, the energy consumption of the device is allocated such that the battery life is maximized while satisfying the constraints.
At the end of the interval, we use the actual energy and activities of the user to update the models.
We also determine the optimal charging time for the device so that critical events are not missed and ambient energy is not wasted. For instance, the battery trajectory on the right shows that we do not need charging at hour one because there is sufficient ambient energy available, while hour 23 needs charging. The AdaEM approach accurately predicts this behavior and schedules charging only for hour 23.

% In addition to optimal energy allocation and battery charging, we must also ensure that the predictions future activity and energy harvest are accurate. The activity patterns of a user and energy harvest can change over time due to changes in personal circumstances or seasons. Therefore, it is critical to ensure that the models used to predict activities and future energy adapt to these changes. To this end, we use the actual energy and activities of the user to update the models at runtime. \textbf{Write 1-2 sentences of conclusion here,}

\section{Dynamic Robust Optimization Approach}

% This section provides a general formulation for maximizing the operating life between recharging operations of a wearable device.

%Maximizing the interval between recharging operations reduces user burden and has the potential to improve the adoption of wearable devices. 

The energy management (EM) problem is challenging due to the following reasons: 1) Energy harvesting from ambient sources is highly stochastic; 2) At the same time, the wearable device should have sufficient energy to fulfill the requirements of target application at all times; and 3) Recharging during periods of important activities (e.g., gait monitoring in a patient with Parkinson's disease) leads to a reduction in the quality of service to the user. Therefore, EM algorithms for wearable devices must account for the future \textit{worst case} uncertainty in the ambient energy harvest and predictions about user activities and energy usage. To overcome these challenges, we formulate the EM decision-making as a \textit{dynamic robust optimization (DyRO)} with constraints on the battery energy and activity accuracy. In what follows, we first describe the DyRO formulation and then provide a light-weight algorithm for real-world deployment.

\subsection{DyRO Formulation}

\vspace{0.75ex}

\noindent {\bf Setup for Decision-making:} Without loss of generality, we consider a set of equal length intervals $\mathcal{T}$ (minutes, hours, etc.) in a given time horizon $T$. The EM decision-making is performed at each discrete interval $t \in \mathcal{T}$ over a fixed time horizon of $T$, such as a day, to account for the repetitive nature of human activities. %Within the horizon, intervals $t \in \mathcal{T}$ are the discrete time periods at which the energy management decisions are made. 
At the beginning of each interval (decision epoch) $t$, the EM algorithm allocates the energy to be consumed in the interval based on the user activities, battery level, and energy harvest. Similarly, any manual charging is scheduled for one or more intervals depending on the energy required to replenish the battery. For example, the EM algorithm can decide to recharge the battery for three consecutive intervals. 

\vspace{0.75ex}

\noindent {\bf Variables and Constraints:} %Using the above setup, 
We define the following variables and constraints to describe the system dynamics.

\vspace{0.5ex}

\textbf{1) Battery recharging indicator $B(t)$:} This binary variable indicates whether recharging is scheduled in a given interval or not. %It is set to 1 if recharging is enabled in an interval, i.e. $B(t) = 1, t \in \mathcal{T}$. Otherwise, it is set to zero. 
At each decision epoch $t \in \mathcal{T}$, the EM algorithm assigns $B(t)$ to 1 (recharge) or 0 for maximizing the duration between recharges while satisfying the constraints.

\vspace{0.5ex}

\textbf{2) Battery energy dynamics:} In each interval $t \in \mathcal{T}$, the battery receives energy from the harvesting sources and a charging source, if the charging is enabled. The harvested energy in an interval $\xi_{t}^H$ is a random variable with uncertainty because environmental conditions can affect the harvested energy. We have a random variable for the energy harvested for each interval in $\mathcal{T}$. Furthermore, the target application will draw energy from the battery to execute the required tasks. The energy used by the application is a decision variable that is computed by the EM algorithm. These dynamics of the battery can be captured as follows: %with the following equation:
\begin{equation}
    E_{t+1}^B = E_{t}^B + \eta \xi_{t}^H + E_t^I B(t) - E_t^c,~~~t \in \mathcal{T}
\end{equation}
where $E_{t}^B$ and $E_{t+1}^B$ are energy at the beginning of the current and next interval respectively, $\eta$ is the EH efficiency, $\xi_{t}^H \in \Xi_{t}^H$ is the realization of harvested energy at interval $t \in \mathcal{T}$, $B(t)$ is the battery charging flag, $E_t^I$ is the charging energy, and $E_{t}^c$ is the energy consumed in the interval $t$.

\vspace{0.5ex}

\textbf{3) Battery energy constraints:} The battery must always maintain a minimum level of charge so that it has sufficient energy to take actions in case of an emergency. For example, if the device detects a patient fall, it must be able to notify a caretaker and call for help. We also ensure that the energy at the end of any horizon $T$ is equal or greater than a target level $E_{target}$ for the next horizon. Similarly, the energy in the battery cannot exceed the design capacity. Therefore, the battery energy constraints can be expressed as:
\begin{equation}
   E_{min} \leq  E_t^B  \leq E_{max}, ~~t \in \mathcal{T},~~~~~E_T^B \geq E_{target}
\end{equation} 
where $E_{min}$ and $E_{max}$ are the min. and max. energy values. %respectively.

\vspace{0.5ex}

\textbf{4) Accuracy constraint:} Without loss of generality, we assume that the device is performing a prediction task for monitoring health of the user. It is critical to ensure that appropriate accuracy is provided to the user and their health providers. Therefore, we impose an accuracy constraint on the application using a threshold $A_{min}$ as follows:
\begin{equation}
    A_t >= A_{min},~~t \in \mathcal{T}
\end{equation}
where $A_t$ is the accuracy in interval $t$.

\vspace{0.5ex}

\textbf{5) Critical activity constraint:} In general, some activities performed by the user have a higher priority for health monitoring than others. For example, in a user with movement disorders, it is critical to monitor for falls or freezing of gait when they are walking. 
% Similarly, if a user has a sleep disorder, it is crucial to monitor their sleep.
We designate a subset of the activities
$\mathcal{A}_c \in \mathcal{A}$ as critical activities, during which recharging cannot be scheduled. 

\vspace{0.75ex}

\noindent \textbf{Optimization Problem:} The primary objective of the EM algorithm is to maximize the utilization of harvested energy and maximize the duration between consecutive charging operations. Suppose the start times of the charging are given by the set $\mathcal{T}_{c} = \{c_1, c_2, \dots, c_m \}$. 
The charging times are a function of the realizations of the harvested energy $\xi_{t}^H, t \in \mathcal{T}$. 
Therefore, the optimization objective is:
\begin{equation} 
    \mathrm{maximize}~~~[ \min \mathbb{E} \{ \Delta \mathcal{T}_{c}(\xi_{1}^H, \cdots, \xi_{T}^H) \}]
\end{equation}
The objective function first takes the first-order difference of the elements in the set $\mathcal{T}_{c}$. The minimum value of the first-order difference gives the shortest time interval between two charging sessions. 
Since our goal is to prolong the operation of the device, we maximize the minimum of the first-order difference of $\mathcal{T}_{c}$. In summary, we can write the overall optimization problem as follows:
\begin{align} \label{eq:objective}
 \mathrm{maximize} &~~~\min \mathbb{E} \{ \Delta \mathcal{T}_{c}(\xi_{1}^H, \cdots, \xi_{T}^H) \}   \\  \label{eq:battery_dynamics}
\text{s. t.}~~~E_{t+1}^B &= E_{t}^B + \eta \xi_{t}^H + E_t^I B(t) - E_t^c,~~~t \in \mathcal{T} \\ \label{eq:battery_target}
  E_{min} &\leq  E_t^B  \leq E_{max}, ~~t \in \mathcal{T},~~E_T^B \geq E_{target} \\ \label{eq:accuracy_constraint}
 A_t & \geq A_{min},~~t \in \mathcal{T} \\ \label{eq:critical_act}
 B(t) &= 0~~\forall t: A_t \in \mathcal{A}_c  
%&&& E_0~\text{is given}
\end{align}

\vspace{0.5ex}

\textbf{Optimization variables:} Solving the above optimization problem involves determining the values of $B(t), E_t^c$, and $E_{t+1}^B$. The battery charging indicator is the primary variable that directly relates to the objective function. 
The EM algorithm must assign a zero or one to each of the intervals in $\mathcal{T}$ while satisfying the constraints. 
$E_t^c$, and $E_{t+1}^B$ are indirect variables which affect the battery dynamics and target application accuracy. Specifically, higher energy consumption $E_t^c$ leads to a faster depletion of the battery, which in turn requires the user to recharge the device sooner.
At the same time, higher energy consumption typically leads to higher accuracy. Therefore, the EM algorithm should minimize the energy consumption of the device while satisfying the accuracy constraints.
Finally, $E_{t+1}^B$ is a function of the energy consumption, harvest, and the charging availability. The goal of the optimization is to ensure that the battery level always stays between $E_{min}$ and $E_{max}$ constraints.

\vspace{0.5ex}

\textbf{Problem complexity:} 
The optimization problem in Equations~\ref{eq:objective}--\ref{eq:critical_act} is an instance of \textit{dynamic robust optimization (DyRO)}~\cite{bertsekas2012dynamic} with a discrete and non-linear objective. In particular, solution to the problem must consider the worst-case uncertainty in the energy harvest to ensure that battery constraints are not violated. Furthermore, as new information becomes available at runtime in the form of actual EH values, the optimization for adaptive EM should change future decisions as a function of the new information.
The uncertainty in the problem, along with a mix of zero/one~($B(t)$) and continuous variables~($E_t^c$ and $E_{t+1}^B$) make the problem computationally hard to solve.
Specifically, obtaining the optimal solution to the general problem formulation requires an exhaustive search, which leads to a time-complexity of $\mathcal{O}(c^{|\mathcal{T}|})$, where $c>1$. If we make the problem convex and use continuous variables for $B(t)$, the time-complexity will be at least $\mathcal{O}(|\mathcal{T}|^3)$ for iterative algorithms.
% objective function in Equation~\ref{eq:objective} is discrete and non-linear due to the minimum operation on $\Delta(\mathcal{T})$. The equality constraint in Equation~\ref{eq:battery_dynamics} is linear while the other constraints are non-linear.
% In terms of the variables, $B(t)$ is a zero-one integer variable while the $E_t^c$, and $E_{t+1}^B$ are continuous variables. 
% Overall, the presence of non-linear objective and constraints, and the zero-one integer variable make the problem computationally hard to solve.
The time-complexity is especially critical for wearable devices since they are energy constrained and repeated calls to computationally expensive algorithms can defeat the purpose of energy management. Therefore, in the next section, we develop a light-weight algorithm that achieves comparable results to the optimal solutions from a solver.

\subsection{AdaEM: a Light-Weight Algorithm for DyRO}\label{sec:solution}

\vspace{0.75ex}

\noindent {\bf Key challenges:} Solving the DyRO problem instance optimally at runtime is infeasible for two key reasons: {\bf 1)} the true values of the uncertain energy harvest for current and future intervals within the decision-making horizon;  and {\bf 2)} wearable devices are constrained by both computing and energy resources needed for solving the DyRO problem.
Therefore, we develop a two-stage approach to design an efficient and practical algorithm referred to as AdaEM to solve the DyRO problem instances for making adaptive EM decisions.

\vspace{0.75ex}

\noindent {\bf Overview of the AdaEM algorithm:} The first stage of AdaEM involves using supervised learning algorithms to learn the distribution and uncertainty of the energy harvest in both current and future intervals. We exploit the historical data and side information about energy harvest for this purpose. For example, the wearable device can record information about energy harvested in the past as features to predict future energy and uncertainty. Similarly, the side information available for wearable devices includes the location (e.g. outdoors vs. indoors), user activity, and current day (e.g., weekday vs. weekend).
This side information can be used to further improve the ML models for energy harvest. 
The energy predictions and uncertainty from the ML models are then used to get a concrete instantiation of the DyRO problem to obtain EM decisions. 

The second stage of AdaEM solves the concrete DyRO problem instance as follows. At the beginning of each decision-making horizon, the predicted energy harvest and uncertainty are used to obtain initial assignments of battery charging and energy consumption allocations, i.e., EM decisions. Subsequently, at runtime, the EM decisions are revised based on the actual values of the harvested energy. This approach ensures that the wearable device takes advantage of any excess harvested energy to extend the battery life or account for shortfall of energy so that the constraints are not violated. In what follows, 
we describe these two stages in more detail. 

\vspace{0.75ex}

\noindent {\bf Energy Harvest Prediction:} The quality of EM decisions depends critically on the accuracy of predicted future energy harvest. In addition to energy harvest prediction, we also need the uncertainty in predictions to reason about the worst-case behavior. Towards this goal, we use the historical energy harvest data to train a ML model using supervised ML algorithms. The input features of the system state $s$, $\phi(s)$, for making predictions include the energy harvest in the past days, hours, and the derivative of the energy harvest in the past hour~\cite{yamin2021online}. We also include side information such as the current location, user activity, and day as part of the feature set $\phi(s)$. The ground-truth energy harvest values $y^*$ are used as supervised data. We have a regression learning problem at hand and any off-the-shelf regression learner can be employed. However, due to the constraints on energy and compute resources in wearables, we need to select a light-weight ML model such as trees, since comparison operations are efficient. 

To choose the energy harvest predictor, we compare the performance of neural networks, linear
regression, and an ensemble of regression trees.
All the learning approaches have a non-negativity constraint for the predictions since energy harvest is always non-negative.
Neural networks and regression trees gave similar accuracy while linear regression showed relatively lower accuracy.
In our specific implementation, we employ an ensemble of regression trees: prediction is the mean and uncertainty is the variance. We note that our final results are not dependent on any particular regression learner. In fact, both neural networks and regression trees for energy prediction lead to similar performance for the overall energy management.

\begin{algorithm}[tb]
\footnotesize
\caption{AdaEM for Charging Optimization}
\label{alg:algorithm}
\textbf{Input}: EH prediction $\hat{\xi}_t^H$ and uncertainty $U$ $\forall t \in \mathcal{T}$, Energy constraints, Accuracy constraints, Energy vs. Accuracy profile\\
\textbf{Output}: Battery recharging indicator $B(t), t \in \mathcal{T}$, Energy consumption, Battery levels
\begin{algorithmic}[1] %[1] enables line numbers
\STATE Set $B(t) = 0$ $\forall t: A_t \in \mathcal{A}_c$
\STATE Set $E_t^c, t \in \mathcal{T}$ such that accuracy is maximized 
% using the energy vs. accuracy profile 
\STATE Calculate $E_{min}^c$ that satisfies the accuracy constraint
\STATE Obtain $E_{t+1}^B, t \in \mathcal{T}$ using EH prediction and assigned $E_t^c$
\WHILE{$\exists t | (E_{t+1}^B < E_{min})$}
% \STATE 
\IF {$\exists t | (E_{t}^c > E_{min}^c)$}
\STATE Reduce $E_{t}^c$ for intervals with higher consumption
\ELSE
\STATE Calculate the energy deficit $\delta E = E_{target} - E_{T+1}^B$
\STATE Calculate number of charging intervals: $\delta E / E^I$
\STATE Set $B(t) = 1$ starting with the first violation
\ENDIF
\STATE Obtain $E_{t+1}^B, t \in \mathcal{T}$ using EH prediction, $E_t^c$, and $B(t)$
\ENDWHILE
\STATE \textbf{return} $B(t), E_t^c, E_{t+1}^B \forall t \in \mathcal{T}$
\end{algorithmic}
\end{algorithm}

% The output of the energy prediction is the expected value of the energy harvest in the future intervals and the uncertainty in the form of the variance. We note that any supervised learning algorithm can be used to predict the future energy and uncertainty. In our specific implementation, we use a neural network to make energy predictions.

\vspace{0.75ex}

\noindent {\bf Charging Optimization Algorithm:} This algorithm takes the predicted energy harvest and uncertainty to determine the charging policy and energy consumption for all intervals in a given decision-making horizon, as shown in Algorithm~\ref{alg:algorithm}.
It is inspired from the water filling algorithm in communication systems~\cite{information_theory} where the goal is to allocate a given power to multiple channels so that the throughput is maximized.
At the beginning of each decision-making horizon, the predicted energy harvest, uncertainty, and the constraints for all intervals are taken as the input. Additionally, it also takes a profile of the energy and accuracy trade-off to ensure that the activity constraints are satisfied. 
The energy and accuracy trade-off are obtained by characterizing the target application on the wearable device. Specifically, we collect activity data
with different sampling rates and design a classifier for each
sampling rate. These datapoints are then used to obtain the
relationship between sampling rate and accuracy~\cite{mirzadeh2020optimal,bhat2019reap}.

After getting the above inputs, the algorithm first initializes the battery charging indicator to zero for intervals where the user performs critical activities. This assignment is based on the history of user's activities. For example, by monitoring a user's activities for a month, we can generate an expected pattern of their activities and use it to set the values of $B(t)$ at the beginning of each horizon. As new data becomes available, the algorithm adjusts the values of $B(t)$ so that any deviations from the expected activity pattern are taken into account.
The algorithm also sets the energy consumption of all intervals to the highest value so that the accuracy is maximized.
Next, using the worst-case EH predictions, we calculate the energy in the battery for each interval in the horizon (Line 4).
If the algorithm detects any violations in the battery levels, it either reduces the energy consumption or enables charging (Lines 5--14). Since our goal is to maximize the battery lifetime, we first reduce the energy consumption of the target application while maintaining the minimum accuracy. If the reduction in energy consumption does not resolve the battery violations, the algorithm calculates the potential deficit in the energy and the number of intervals required to complete the charging.
Finally, it sets the battery charging indicators for the required number of intervals to one, starting with the first violation. 

The initial values of the energy consumption and battery charging are based on the EH predictions. At runtime, the actual energy is typically different from the predicted values. Therefore, as data about the actual EH becomes available, we re-run Algorithm~\ref{alg:algorithm} with a shorter horizon to adjust the energy consumption and battery charging.
For instance, if excess energy is harvested in an interval, we can either increase the energy consumption in the following intervals to improve the accuracy or reduce the time spent in charging.
The time complexity of the algorithm is $\mathcal{O}(|\mathcal{T}|^2)$ in the worst-case when the while loop runs $|\mathcal{T}|$ times, where  $|\mathcal{T}|$ stands for the number of decision-making steps.

\section{Experiments and Results}

In this section, we present the experimental results for the proposed AdaEM algorithm along different dimensions.

\subsection{Experimental Setup}
\noindent\textbf{Wearable device:} We employ a prototype based on the Texas Instruments (TI) CC2652R micro-controller (MCU)~\cite{ticc2652} as the primary wearable device in our experimental validation. The TI-CC2652 integrates a low-power ARM Cortex M4 core and interfaces such as SPI and I2C for sensors.
In addition to the MCU, we include a SP3-37 flexible PV cell~\cite{sp3-37} for harvesting energy from ambient light and a piezoelectric harvester for motion EH.  
The wearable device incorporates a GMB 031009~\cite{031009} Lithium polymer battery with a capacity of 12 mAH~(160 J) as the energy storage element.

% \noindent\textbf{Driver Application:} We use activity monitoring as the driver application for validating our energy management approach since it has wide applicability to a variety of health applications~\cite{bao2004activity}.
% Specifically, we use the inertial motion units on our wearable device to monitor the motion of the user's legs and identify activities such as walking, sitting, standing, and jumping.

\vspace{0.5mm}
\noindent\textbf{Activity Dataset:}
We use the American Time Use Survey~(ATUS)~\cite{amtus} to obtain the user activity data. The ATUS, conducted by the U.S. Department of Labor, contains typical activity data for approximately 10,000 users.
We pre-process the ATUS data to obtain five main activity categories: \textit{\{Sleep, work, exercise, leisure, and others\}}. The wearable device is then used to identify these activities using the sensors available on the device. Of these activities, we choose `exercise' as the critical activity. This means that battery recharging must not be scheduled in time intervals when the user is exercising.

\vspace{0.5mm}
\noindent\textbf{Energy Harvesting Dataset:} The potential energy available from the motion harvesters is a function of the user motion during the activity. To this end, we set the activity intensity for legs and hands for each activity. Starting with a baseline intensity of 1 for exercise, we set the intensity for the `active' activity as 0.5, while the intensity for sleeping is zero. 
The activity intensities are then combined with a baseline power of 13~$\upmu$W~\cite{tuncel2020towards} available from the piezoelectric harvesters.
Similarly, the energy available from light is a function of the location. To illustrate the performance of AdaEM clearly, we use outdoor light intensity throughout the day. Specifically, we utilize five years~(2016--2020) of solar irradiation measured by National Renewable Energy Laboratory (NREL)~\cite{NREL} in Golden, Colorado. The irradiance is combined with the time of the day to calculate the light energy using the area and I-V characteristics of the PV cell~\cite{sp3-37}.

\vspace{0.5mm}
\noindent\textbf{Experiment Parameters:} 
% The parameters needed for our experimental validation include the length of the horizon, intervals, and the energy constraints.
We set the finite horizon to be 24 hours, since EH and user activities typically follow a daily pattern~\cite{huynh2008discovery}. The length of each interval is set to one hour while noting that other intervals also work well with AdaEM.
Finally, we set the minimum energy constraint to 10\% of the battery capacity and the target energy constraint to 60\% of the battery capacity.

\vspace{0.5mm}
\noindent\textbf{Optimal Solution:} We obtain the optimal solution using the CVX package~\cite{cvx} in Matlab. Specifically, we make the problem convex by minimizing the sum of $B(t)$. This is feasible since the optimization is done over a finite horizon. As a result, minimizing the sum of $B(t)$ maximizes the operating time of the device in each horizon. 
We make the problem convex to avoid the prohibitive cost of an exhaustive search.
The CVX implementation is run on a server consisting of 32 Intel\textsuperscript{\textregistered} Xeon\textsuperscript{\textregistered} Gold 6226R cores with 192 GB of memory.
As noted earlier, the optimal solution uses the actual energy harvest values with no uncertainty.

\begin{figure}[b] 
    \centering
    \includegraphics[width=0.83\linewidth]{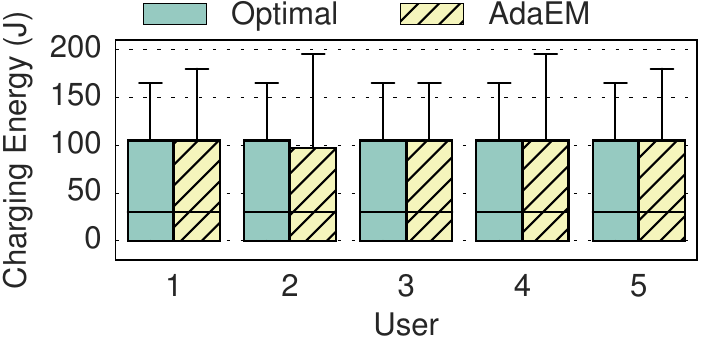}
    \caption{Energy used for recharging by the optimal and proposed approaches for each month of the year (2016--2020) under ideal EH predictions}
     \label{fig:charging_actual}
\end{figure}

\vspace{0.5mm}
\noindent\textbf{On-demand Baseline Algorithm:} We also implement a baseline that aims for a fixed accuracy and begins charging whenever the battery drops below $E_{min}$. The charging continues until the battery reaches $E_{target}$. We refer to this charging policy as the on-demand baseline algorithm since it enables charging whenever the energy drops below $E_{min}$ and more energy is needed.
This policy reflects the behavior of most wearable and smartphone users.
As a result, it provides a useful baseline to evaluate the efficacy of AdaEM.

% When compared to AdaEM, the new baseline policy results in a much higher degree of battery constraint violations (Fig.~1).
% More importantly, the baseline does not account for critical activities and has insufficient energy for critical activities. In contrast, AdaEM uses activity predictions to ensure that critical activities are detected while satisfying all the battery constraints. 

\subsection{Energy Management with Ideal Predictions}
We start with an ablation study where we provide the actual EH values to the adaptive charging algorithm. Results obtained using the actual EH give us the upper bound on the performance of the proposed charging optimization algorithm.
Figure~\ref{fig:charging_actual} shows a distribution of the daily charging energy required for five distinct users over a period of five years.
We see that the median values of the charging energy for AdaEM are within 1\% to the optimal solution. Furthermore, the range and distribution of the charging energies are close to the optimal solution.
This shows that AdaEM is able to accurately determine the charging intervals with significantly lower computations than the optimal algorithm.

\begin{figure}[t] 
    \centering
    \includegraphics[width=0.83\linewidth]{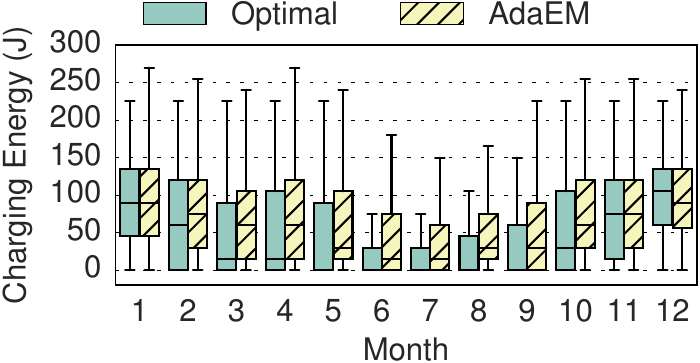}
    \caption{Energy used for recharging by the optimal and proposed approaches for each month of the year (2016--2020) under uncertain EH predictions}
    \label{fig:charging_energy_per_month}
\end{figure}
\subsection{Energy Management with Uncertainty in EH}
% Real world usage of the energy management algorithm must account for day to day variations in the harvested energy as well as the error in prediction of the harvested energy. 
In this section, we analyze the performance of AdaEM with uncertain energy harvest predictions.
To this end, we first predict the available energy harvest at the beginning of each day using the trained regression learner.
This energy prediction is used by Algorithm~\ref{alg:algorithm} to assign the energy consumption and charging policy for each interval of the day while satisfying all the constraints. 
Since the actual energy can deviate from the predicted values, the algorithm is re-run at the beginning of each interval to account for these differences.
Figure~\ref{fig:charging_energy_per_month} shows the comparison of the charging energy used by the proposed approach and the optimal solution. 
The optimal solution is computed offline and it uses the actual values of harvested energy.
We see that, in general, the energy used by the proposed approach is close to the optimal solution for most months. 
% In particular, the median value of the charging energy matches the optimal value for the winter months.
A higher variation is observed for summer months, where the distribution of AdaEM is wider than the optimal solution. This is because in summer months the algorithm generally expects a higher solar energy harvest. Therefore, when it encounters cloudy days with a lower solar energy harvest, the prediction error increases, which, in turn, causes the device to use more energy for charging.

\begin{figure}[t] 
    \centering
    \includegraphics[width=0.83\linewidth]{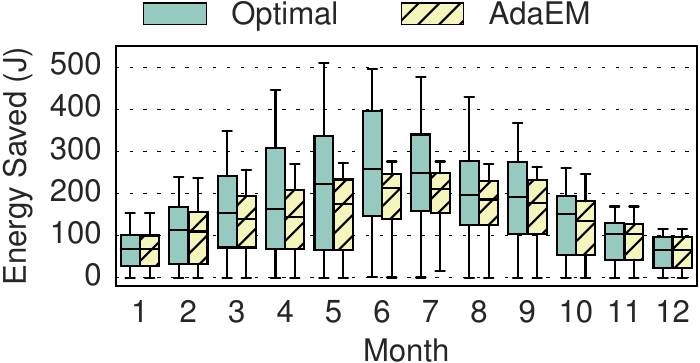}
    \caption{Energy saved by using energy harvesting and sustainable allocation by the optimal and proposed approaches for each month of the year (2016--2020)}
    \label{fig:energy_savings}
\end{figure}

\vspace{1mm}
\noindent\textbf{Energy Savings:} One of the primary advantages of sustainable operation and optimal charging is the savings in the total energy consumption through use of EH. Figure~\ref{fig:energy_savings} shows the distribution of energy savings achieved by the proposed approach when compared to the optimal solution for each month of the year over a period of five years. The savings here represent the additional energy that would be needed if EH is not used.
As expected, the energy savings are higher in summer months due to higher energy availability. The optimal solution is able to achieve higher savings than AdaEM because it has full information about the future energy harvest. 
At the same time, the proposed approach achieves median savings that are within 5\% of the optimal values.
We also see that both approaches achieve significant savings even in winter months when the harvested energy is lower.
% This shows that the proposed approach is effective in achieving sustainable operation in energy constrained months.

\begin{figure}[t] 
    \centering
    \includegraphics[width=0.8\linewidth]{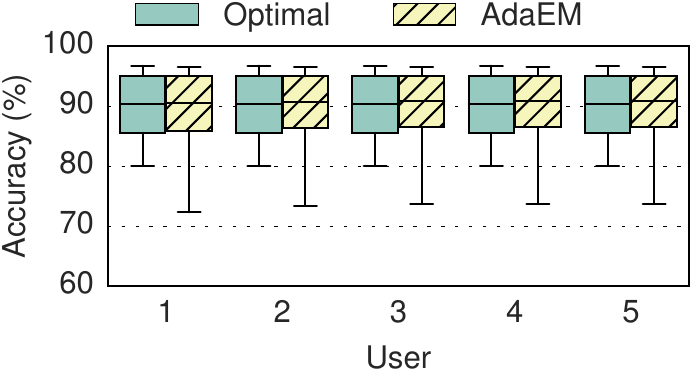}
    \caption{Average daily accuracy of activity recognition for five users over a period of five years}
    \label{fig:cac_acc}
\end{figure}

\vspace{1mm}
\noindent\textbf{Effect of the Activity Constraints:}
Next, we analyze the impact of the critical activity constraint on the accuracy of the application. We choose five users with varying lengths of the critical activity constraint. Specifically, user 1 has the longest critical activity constraint with six intervals while user 5 has a critical activity constraint with two intervals.
The distribution of average daily accuracy for the five users is shown in Figure~\ref{fig:cac_acc}. 
The minimum accuracy constraint in the figure is 90\%.
The optimal solution meets the minimum accuracy constraint for all the users. 
On the other hand, AdaEM has few outliers where the accuracy constraint is not met. This happens on days with limited energy harvest and the device is unable to recharge the battery in time. As a result, the energy consumption has to be reduced, which leads to a reduction in the accuracy.
We observe that the number of outliers decreases as the length of the critical activity constraint decreases (user 1 to user 5). 
This is because a lower number of critical activity intervals gives more opportunity for the device to recharge. In summary, these experimental results show that AdaEM efficiently enables sustainable operation of wearable devices for health monitoring.

%  whenever the actual energy deviates significantly from the predicted values

\begin{figure}[t] 
    \centering
    \includegraphics[width=1\linewidth]{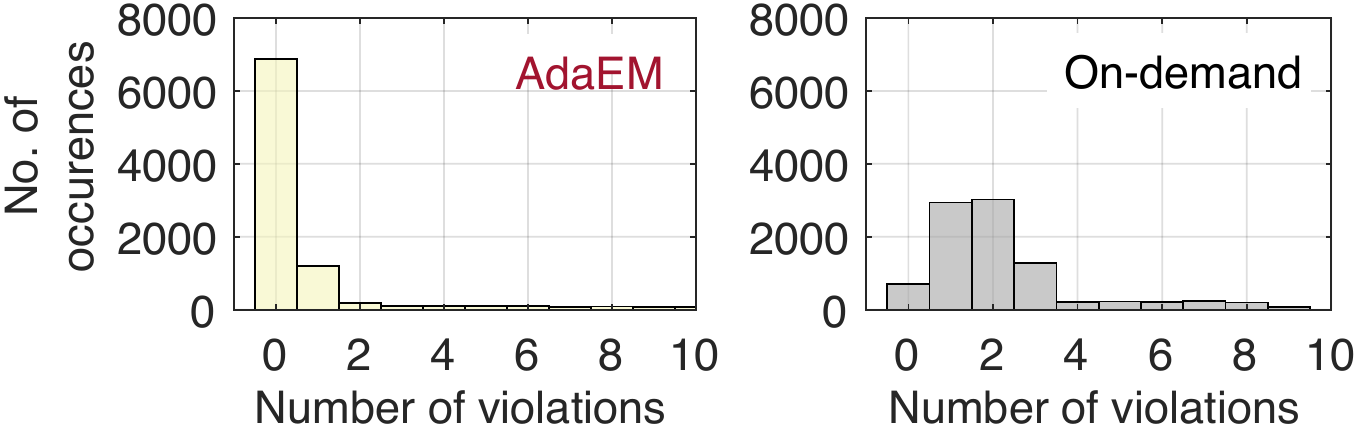}
    \caption{Constraint violations for AdaEM and the On-demand baseline algorithm}
    \label{fig:violation}
\end{figure}

\subsection{Comparison to the Baseline Algorithm}
In this section we compare AdaEM with the baseline algorithm. Recall that the On-demand algorithm begins charging whenever the battery level drops below $E_{min}$.
This approach is not suitable for sustainable health monitoring because of two key limitations: 1) the baseline does not account for critical activity constraints.
Consequently, the device may have insufficient energy to provide high accuracy for critical activities, and 2) 
Unlike AdaEM, the On-demand algorithm is reactive in nature and does not plan for satisfying the minimum and target battery energy constraints. As a result, the quality of service to the application can suffer significantly.

To illustrate the above limitations, we compare the energy constraint violations for AdaEM and the On-demand algorithm in Figure~\ref{fig:violation}. The optimal solution is not included in the figure because it does not have any constraint violations due to the knowledge of exact energy harvest and consumption.
The figure shows that the On-demand algorithm suffers a significantly higher degree of constraint violations when compared to AdaEM. Specifically, it has more than 3000 occurrences of two or more violations over five years of evaluation data for five users. In contrast, AdaEM has negligible number of instances with two or more constraint violations.
AdaEM is able reduce the number of violations and satisfy the critical activity constraints by leveraging future EH and user activity predictions in making EM decisions. 
In summary, the comparison with the baseline shows the effectiveness of the proposed AdaEM approach in achieving sustainable operation for wearable health monitoring devices.

\subsection{Comparison to Energy Neutral Methods}
\begin{figure}[t] 
    \centering
    \includegraphics[width=0.85\linewidth]{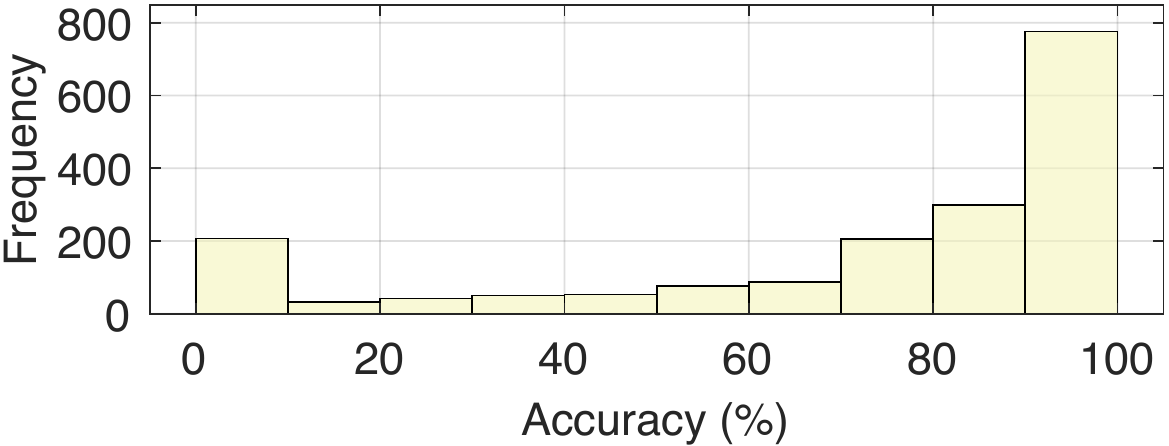}
    \caption{Average daily accuracy of activity recognition for five users over a period of five years}
    \label{fig:cvx_acc}
\end{figure}

Energy neutral methods that rely only on harvested energy have been popular for wearable devices~\cite{kansal2007power}. However, as we noted before, energy neutral methods rely on duty cycling the device, which can lead to reduced performance.
To illustrate this, we implemented an optimal energy neutral approach that uses the actual harvested energy from 2016 to 2020. The energy neutral approach allocates the device energy consumption such that the total energy consumed in a day is equal to the harvested energy. Figure~\ref{fig:cvx_acc} shows a histogram of the accuracies achieved over five years. As we can see, a significant number of days have close to zero accuracy.
As a result, the user is unable to monitor their health parameters, which can lead to adverse effects.
This shows the benefits of using AdaEM to balance the accuracy requirements of the user while utilizing harvested energy.

\subsection{Implementation Overhead}
We characterize the overhead of Algorithm~\ref{alg:algorithm} by implementing it on the TI-CC2652 MCU. 
% Specifically, we implement the algorithm in C to measure the execution time and power consumption when making energy management decisions.
We observe that the algorithm takes 3~ms to run at a power consumption of 11~mW. This amounts to 33~$\upmu$J of energy. Since the algorithm runs once every hour, the total energy consumption over a day is 792~$\upmu$J, which is less than 0.005\% of the battery capacity. This shows that AdaEM achieves sustainable operation with negligible overhead.

\section{Social Impact}
The social impact potential of the proposed AdaEM approach is two-fold. First, AdaEM can catalyze widespread adoption of wearable devices by enabling sustainable operation with high quality of service. 
Specifically, our holistic approach of combining energy harvesting and management with flexible hybrid electronics leads to sustainable operation with user comfort.
This will, in turn, have significant advantages in public health and healthcare expenditure.

Second, it is projected that there will be billions of internet of things (IoT) devices by 2050. Integrating AdaEM into the IoT devices will help in reducing the overall energy footprint through sustainable operation.
For instance, even savings of 100~J per day for each device represents a decrease of 3*10$^7$ units of electricity per year.
This energy is equivalent to the consumption of 20,000 U.S. households, as per the~\citet{eia}.

% More details on the potential broader social impacts are presented in the Appendix.

% The social impact potential of the proposed AdaEM approach is two-fold. First, AdaEM can enable widespread adoption of wearable devices. 
% Specifically, our holistic approach of combining energy harvesting and management with flexible hybrid electronics leads to sustainable operation with user comfort. 
% This will, in turn, have significant advantages in public health and healthcare expenditure.

% Second, it is projected that there will be 24 billion of internet of things (IoT) devices by 2050~\cite{ericsson}. Integrating AdaEM into the IoT devices will help in reducing the overall energy footprint through sustainable operation.
% For instance, even savings of 100~J per day for each device represents a decrease of 3*10$^7$ units of electricity per year.
% This energy is equivalent to the consumption of 20,000 U.S. households, as per the Energy Information Administration~\cite{eia}.

\section{Conclusion and Future Work}
Wearable devices present a tremendous potential to change the landscape in health monitoring. 
However, their widespread adoption has been hindered by frequent recharging requirements. This paper presented the novel problem of \textit{adaptive energy management} towards sustainable operation of wearable devices. Starting with a dynamic robust optimization, we presented a light-weight solution that uses ML predictions of energy harvest to optimize the charging of the wearable device.
Using real-world data and a wearable prototype, we showed that AdaEM achieves solutions that are within 5\% of the optimal with less than 0.005\% overhead. 
Our immediate future work includes deploying the proposed solution in the field by performing user studies.

The dynamic robust optimization formulation presented in this paper has applications in a number of fields including agricultural monitoring, wide area sensing, and defense. Even in the area of mobile health, accounting for sudden changes in user activity remains a challenge. For instance, conditions such as freezing of gait, cardiac arrest, or falls are difficult to predict using the history of past activities. 
The challenge in the activity prediction further complicates the energy management problem since the algorithm must account for these sudden changes in an energy-efficient manner. 
We hope that this paper inspires the artificial intelligence community to develop new algorithms in this exciting area of self-sustainable wearable devices.
% .

\section*{Acknowledgments}
This work was supported in part by the Washington State University New Faculty Seed Grant to Ganapati Bhat.

% Use \bibliography{yourbibfile} instead or the References section will not appear in your paper
% \bibliography{references/wearable_iot,references/embedded_refs,references/flexible,references/health_refs}

\begin{thebibliography}{42}
\providecommand{\natexlab}[1]{#1}

\bibitem[{Andreas and Stoffel(1981)}]{NREL}
Andreas, A.; and Stoffel, T. 1981.
\newblock {NREL Solar Radiation Research Laboratory (SRRL): Baseline
  Measurement System (BMS); Golden, Colorado (Data); NREL Report No.
  DA-5500-56488}.
\newblock Accessed 3/28/2021.

\bibitem[{Barrera et~al.(2020)Barrera, Reina, Mat{\'e}, and
  Trujillo}]{barrera2020solar}
Barrera, J.~M.; Reina, A.; Mat{\'e}, A.; and Trujillo, J.~C. 2020.
\newblock Solar Energy Prediction Model Based on Artificial Neural Networks and
  Open Data.
\newblock \emph{Sustainability}, 12(17): 6915.

\bibitem[{Bertsekas(2012)}]{bertsekas2012dynamic}
Bertsekas, D. 2012.
\newblock \emph{{Dynamic Programming and Optimal Control: Volume I}}, volume~1.
\newblock Athena scientific.

\bibitem[{Bhat et~al.(2019)Bhat, Bagewadi, Lee, and Ogras}]{bhat2019reap}
Bhat, G.; Bagewadi, K.; Lee, H.~G.; and Ogras, U.~Y. 2019.
\newblock {REAP: Runtime Energy-Accuracy Optimization for Energy Harvesting IoT
  Devices}.
\newblock In \emph{Proc. of Annual Design Autom. Conf.}, 171:1--171:6.

\bibitem[{Bhat et~al.(2017)}]{bhat2017near}
Bhat, G.; et~al. 2017.
\newblock {Near-Optimal Energy Allocation for Self-Powered Wearable Systems}.
\newblock In \emph{Proc. Int. Conf. on Comput.-Aided Design}, 368--375.

\bibitem[{Chong et~al.(2019)Chong, Ismail, Ko, and Lee}]{chong2019energy}
Chong, Y.-W.; Ismail, W.; Ko, K.; and Lee, C.-Y. 2019.
\newblock Energy harvesting for wearable devices: A review.
\newblock \emph{IEEE Sensors Journal}, 19(20): 9047--9062.

\bibitem[{Cover and Thomas(2012)}]{information_theory}
Cover, T.~M.; and Thomas, J.~A. 2012.
\newblock \emph{{Elements of Information Theory}}.
\newblock Wiley.

\bibitem[{Daneault(2018)}]{daneault2018could}
Daneault, J.-F. 2018.
\newblock {Could Wearable and Mobile Technology Improve the Management of
  Essential Tremor?}
\newblock \emph{Frontiers in neurology}, 9: 257:1--257:8.

\bibitem[{Dempsey(2015)}]{dempsey2015teardown}
Dempsey, P. 2015.
\newblock The teardown: Apple Watch.
\newblock \emph{Engineering \& Technology}, 10(6): 88--89.

\bibitem[{Dhillon et~al.(2020)Dhillon, Madhu, Kaur, and
  Singh}]{dhillon2020solar}
Dhillon, S.; Madhu, C.; Kaur, D.; and Singh, S. 2020.
\newblock {A Solar Energy Forecast Model using Neural Networks: Application for
  Prediction of Power for Wireless Sensor Networks in Precision Agriculture}.
\newblock \emph{Wireless Personal Comm.}, 1--20.

\bibitem[{Dieleman et~al.(2020)Dieleman, Cao, Chapin, Chen, Li, Liu, Horst,
  Kaldjian, Matyasz, Scott et~al.}]{dieleman2020us}
Dieleman, J.~L.; Cao, J.; Chapin, A.; Chen, C.; Li, Z.; Liu, A.; Horst, C.;
  Kaldjian, A.; Matyasz, T.; Scott, K.~W.; et~al. 2020.
\newblock US health care spending by payer and health condition, 1996-2016.
\newblock \emph{Jama}, 323(9): 863--884.

\bibitem[{Dorsey et~al.(2016)Dorsey, Vlaanderen, Engelen, Kieburtz, Zhu,
  Biglan, Faber, and Bloem}]{dorsey2016moving}
Dorsey, E.~R.; Vlaanderen, F.~P.; Engelen, L.~J.; Kieburtz, K.; Zhu, W.;
  Biglan, K.~M.; Faber, M.~J.; and Bloem, B.~R. 2016.
\newblock {Moving Parkinson Care to the Home}.
\newblock \emph{Movement Disorders}, 31(9): 1258--1262.

\bibitem[{{Energy Information Administration}(2021)}]{eia}
{Energy Information Administration}. 2021.
\newblock {How much electricity does an American home use?}
\newblock \url{https://www.eia.gov/tools/faqs/faq.php?id=97&t=3}, accessed 11
  September 2021.

\bibitem[{Espay et~al.(2016)}]{espay2016technology}
Espay, A.~J.; et~al. 2016.
\newblock {Technology in Parkinson's Disease: Challenges and Opportunities}.
\newblock \emph{Movt. Disorders}, 31(9): 1272--1282.

\bibitem[{{FlexSolarCells}(2013)}]{sp3-37}
{FlexSolarCells}. 2013.
\newblock {SP3-37 Datasheet}.
\newblock \url{https://bit.ly/3dcJ0lK}, accessed 3/28/2021.

\bibitem[{Geissdoerfer et~al.(2019)Geissdoerfer, Jurdak, Kusy, and
  Zimmerling}]{geissdoerfer2019getting}
Geissdoerfer, K.; Jurdak, R.; Kusy, B.; and Zimmerling, M. 2019.
\newblock {Getting more Out of Energy-Harvesting Systems: Energy Management
  under Time-Varying Utility with Preact}.
\newblock In \emph{Proceedings of the 18th International Conference on
  Information Processing in Sensor Networks}, 109--120.

\bibitem[{{GMB}(2009)}]{031009}
{GMB}. 2009.
\newblock 031009 datasheet.
\newblock \url{http://www.gmbattery.com/Datasheet/LIPO/LIPO-031009-12mAh.pdf},
  accessed 5 August 2017.

\bibitem[{Grant and Boyd(2014)}]{cvx}
Grant, M.; and Boyd, S. 2014.
\newblock {{CVX}: Matlab Software for Disciplined Convex Programming, Version
  2.1}.
\newblock \url{http://cvxr.com/cvx}, accessed 5 August 2017.

\bibitem[{Huang, Huang, and Cheng(2011)}]{huang2011robust}
Huang, T.-C.; Huang, J.-L.; and Cheng, K.-T. 2011.
\newblock {Robust Circuit Design for Flexible Electronics}.
\newblock \emph{IEEE Design and Test of Computers}, 28(6): 8--15.

\bibitem[{Huynh, Fritz, and Schiele(2008)}]{huynh2008discovery}
Huynh, T.; Fritz, M.; and Schiele, B. 2008.
\newblock {Discovery of Activity Patterns using Topic Models}.
\newblock In \emph{Proc. of the 10th Int. Conf. on Ubiquitous computing},
  10--19.

\bibitem[{Hyland et~al.(2016)Hyland, Hunter, Liu, Veety, and
  Vashaee}]{hyland2016wearable}
Hyland, M.; Hunter, H.; Liu, J.; Veety, E.; and Vashaee, D. 2016.
\newblock Wearable thermoelectric generators for human body heat harvesting.
\newblock \emph{Applied Energy}, 182: 518--524.

\bibitem[{Johansson, Malmgren, and Murphy(2018)}]{johansson2018wearable}
Johansson, D.; Malmgren, K.; and Murphy, M. 2018.
\newblock {Wearable Sensors for Clinical Applications in Epilepsy,
  Parkinson’s Disease, and Stroke: A Mixed-Methods Systematic Review}.
\newblock \emph{Jrnl. of neurology}, 1--13.

\bibitem[{Kansal et~al.(2007)Kansal, Hsu, Zahedi, and
  Srivastava}]{kansal2007power}
Kansal, A.; Hsu, J.; Zahedi, S.; and Srivastava, M.~B. 2007.
\newblock {Power Management in Energy Harvesting Sensor Networks}.
\newblock \emph{ACM Trans. Embedd. Comput. Syst.}, 6(4): 32.

\bibitem[{Katz and Huang(2009)}]{katz2009thin}
Katz, H.~E.; and Huang, J. 2009.
\newblock {Thin-Film Organic Electronic Devices}.
\newblock \emph{Annual Review of Materials Research}, 39: 71--92.

\bibitem[{Khalifa et~al.(2017)Khalifa, Lan, Hassan, Seneviratne, and
  Das}]{khalifa2017harke}
Khalifa, S.; Lan, G.; Hassan, M.; Seneviratne, A.; and Das, S.~K. 2017.
\newblock Harke: Human activity recognition from kinetic energy harvesting data
  in wearable devices.
\newblock \emph{IEEE Transactions on Mobile Computing}, 17(6): 1353--1368.

\bibitem[{Khan et~al.(2016)}]{khan2016flexible}
Khan, Y.; et~al. 2016.
\newblock {Flexible Hybrid Electronics: Direct Interfacing of Soft and Hard
  Electronics for Wearable Health Monitoring}.
\newblock \emph{Advanced Functional Materials}, 26(47): 8764--8775.

\bibitem[{{LIVMOR}(2021)}]{livmor}
{LIVMOR}. 2021.
\newblock {LIVMOR Halo+}.
\newblock \url{https://www.livmor.com}, accessed 9/7/2021.

\bibitem[{Mirzadeh and Ghasemzadeh(2020)}]{mirzadeh2020optimal}
Mirzadeh, S.~I.; and Ghasemzadeh, H. 2020.
\newblock Optimal Policy for Deployment of Machine Learning Models on
  Energy-Bounded Systems.
\newblock In \emph{Proceedings of the Twenty-Ninth International Joint
  Conference on Artificial Intelligence}.

\bibitem[{Mitcheson et~al.(2008)Mitcheson, Yeatman, Rao, Holmes, and
  Green}]{mitcheson2008energy}
Mitcheson, P.~D.; Yeatman, E.~M.; Rao, G.~K.; Holmes, A.~S.; and Green, T.~C.
  2008.
\newblock Energy harvesting from human and machine motion for wireless
  electronic devices.
\newblock \emph{Proceedings of the IEEE}, 96(9): 1457--1486.

\bibitem[{Ozanne et~al.(2018)Ozanne, Johansson, H{\"a}llgren~Graneheim,
  Malmgren, Bergquist, and Alt~Murphy}]{ozanne2018wearables}
Ozanne, A.; Johansson, D.; H{\"a}llgren~Graneheim, U.; Malmgren, K.; Bergquist,
  F.; and Alt~Murphy, M. 2018.
\newblock {Wearables in Epilepsy and Parkinson's disease—A Focus Group
  Study}.
\newblock \emph{Acta Neurologica Scandinavica}, 137(2): 188--194.

\bibitem[{Park et~al.(2017)}]{park2017pv}
Park, J.; et~al. 2017.
\newblock {Flexible PV-cell Modeling for Energy Harvesting in Wearable IoT
  Applications}.
\newblock \emph{ACM Trans. Embedd. Comput. Syst.}

\bibitem[{Park et~al.(2009)Park, Kim, Stryakhilev, Lee, An, Pyo, Lee, Mo, Jin,
  and Chung}]{park2009flexible}
Park, J.-S.; Kim, T.-W.; Stryakhilev, D.; Lee, J.-S.; An, S.-G.; Pyo, Y.-S.;
  Lee, D.-B.; Mo, Y.~G.; Jin, D.-U.; and Chung, H.~K. 2009.
\newblock {Flexible Full Color Organic Light-Emitting Diode Display on
  Polyimide Plastic Substrate Driven by Amorphous Indium Gallium Zinc Oxide
  Thin-Film Transistors}.
\newblock \emph{Applied Physics Letters}, 95(1): 013503.

\bibitem[{Poliks et~al.(2016)Poliks, Turner, Ghose, Jin, Garg, Gui, Arias,
  Kahn, Schadt, and Egitto}]{poliks2016wearable}
Poliks, M.; Turner, J.; Ghose, K.; Jin, Z.; Garg, M.; Gui, Q.; Arias, A.; Kahn,
  Y.; Schadt, M.; and Egitto, F. 2016.
\newblock A Wearable Flexible Hybrid Electronics ECG Monitor.
\newblock In \emph{Electronic Components and Tech. Conf.}, 1623--1631. IEEE.

\bibitem[{Shresthamali, Kondo, and Nakamura(2017)}]{shresthamali2017adaptive}
Shresthamali, S.; Kondo, M.; and Nakamura, H. 2017.
\newblock Adaptive power management in solar energy harvesting sensor node
  using reinforcement learning.
\newblock \emph{ACM Transactions on Embedded Computing Systems (TECS)}, 16(5s):
  1--21.

\bibitem[{{Texas Instruments Inc.}(2018)}]{ticc2652}
{Texas Instruments Inc.} 2018.
\newblock {CC2652R Microcontroller}.
\newblock [Online] \url{https://www.ti.com/product/CC2652R}, accessed 1
  November 2020.

\bibitem[{Tuncel et~al.(2021)Tuncel, Bhat, Park, and Ogras}]{9519688}
Tuncel, Y.; Bhat, G.; Park, J.; and Ogras, U. 2021.
\newblock ECO: Enabling Energy-Neutral IoT Devices through Runtime Allocation
  of Harvested Energy.
\newblock \emph{IEEE Internet of Things Journal}, 1--1.

\bibitem[{Tuncel et~al.(2020)}]{tuncel2020towards}
Tuncel, Y.; et~al. 2020.
\newblock {Towards Wearable Piezoelectric Energy harvesting: Modeling and
  Experimental Validation}.
\newblock In \emph{Proceedings of the ACM/IEEE International Symposium on Low
  Power Electronics and Design}, 55--60.

\bibitem[{{US Department of Labor}(2015)}]{amtus}
{US Department of Labor}. 2015.
\newblock {American Time Use Survey}.
\newblock \url{https://www.bls.gov/tus/}, accessed 25 July 2017.

\bibitem[{Valenzuela(2008)}]{valenzuela2008energy}
Valenzuela, A. 2008.
\newblock {Energy Harvesting for No-Power Embedded Systems}.
\newblock \url{https://bit.ly/3fnA6Vm}, accessed 3/28/2021.

\bibitem[{Vigorito et~al.(2007)}]{vigorito2007adaptive}
Vigorito, C.~M.; et~al. 2007.
\newblock {Adaptive Control of Duty Cycling in Energy-Harvesting Wireless
  Sensor Networks}.
\newblock In \emph{Proc. Conf. on Sensor, Mesh and Ad Hoc Comm. and Networks},
  21--30.

\bibitem[{Yamin and Bhat(2021)}]{yamin2021online}
Yamin, N.; and Bhat, G. 2021.
\newblock {Online Solar Energy Prediction for Energy-Harvesting Internet of
  Things Devices}.
\newblock In \emph{2021 IEEE/ACM International Symposium on Low Power
  Electronics and Design (ISLPED)}, 1--6.

\bibitem[{Yamin, Bhat, and Doppa(2022)}]{yamin2021diet}
Yamin, N.; Bhat, G.; and Doppa, J.~R. 2022.
\newblock {DIET: A Dynamic Energy Management Approach for Wearable Health
  Monitoring Devices}.
\newblock In \emph{2022 Design, Automation \& Test in Europe Conference \&
  Exhibition (DATE)}, 1--6.

\end{thebibliography}
\newpage
\appendix
\section*{Technology Enablers for Wearable Devices}

\noindent\textbf{Flexible hybrid electronics:}
Early implementations wearable devices included rigid components on a rigid printed circuit board. However, this form factor is not suitable for long term usage because it does not conform to the shape of the body, making it uncomfortable. 
Indeed, recent user studies have shown that comfort is one of the most common reasons why users stop using rigid wearable devices~\cite{ozanne2018wearables}.
Flexible electronics can enable comfortable wearables through use of materials that are fully bendable and rollable.
However, the performance capabilities of fully flexible electronics are low compared to conventional CMOS devices.
For instance, the feature sizes in flexible electronics are limited to $\mu$m range and processor frequencies of 10~MHz~\cite{huang2011robust,katz2009thin}.
To address these limitations, recent research has proposed flexible hybrid electronics~(FHE)~\cite{khan2016flexible}.
FHE uses rigid components on a flexible substrate to implement a device that conforms to the shape of the user's body. Examples of rigid components include processors and sensor while flexible components include antennas, batteries, and displays. By using rigid components for processing and sensing, FHE devices are able to utilize the performance advantages provided by CMOS technology and form factor advantages of the substrate.
Examples of successful implementations of FHE include the devices described in~\cite{poliks2016wearable, khan2016flexible}

\noindent\textbf{Thermoelectric Energy Generators:}
Thermoelectic energy generators~(TEG) uses the temperature difference between to side of the device to convert heat energy to electrical energy. The human body typically maintains an average body temperature of about 37~$^\circ$C by dissipating heat.
% As part of this process, the human body continuously dissipates heat in the order of 100 W~\citep{riemer2010biomechanical}.
TEGs can tap into this energy and use it to charge batteries on a wearable device.
Currently available TEGs have about 3\% efficiency due to high contact and thermal resistance~\cite{hyland2016wearable}. Despite this, TEGs are a promising solution for sustainable operation since they provide a near-constant supply of energy to the wearable device.

\section*{Additional Experimental Results}
\subsubsection*{Optimal Solution:} We obtain the optimal solution using the CVX package~\cite{cvx} in Matlab. Specifically, we make the problem convex by optimizing the sum of the $B(t)$ variable. This relaxation to the convex form is feasible since the optimization is done over a finite horizon. As a result, minimizing the sum of $B(t)$ maximizes the operating time of the device in each horizon. 
We perform the relaxation to avoid the prohibitive cost of an exhaustive search.
The CVX implementation is run on a server consisting of 32 Intel(R) Xeon(R) Gold 6226R with 192 GB of memory.
As noted earlier, the optimal solution uses the actual energy harvesting values with no uncertainty.

\subsubsection*{Effect of the Minimum Accuracy Constraint:} 
\begin{figure}[t] 
	\centering
	\includegraphics[width=0.8\linewidth]{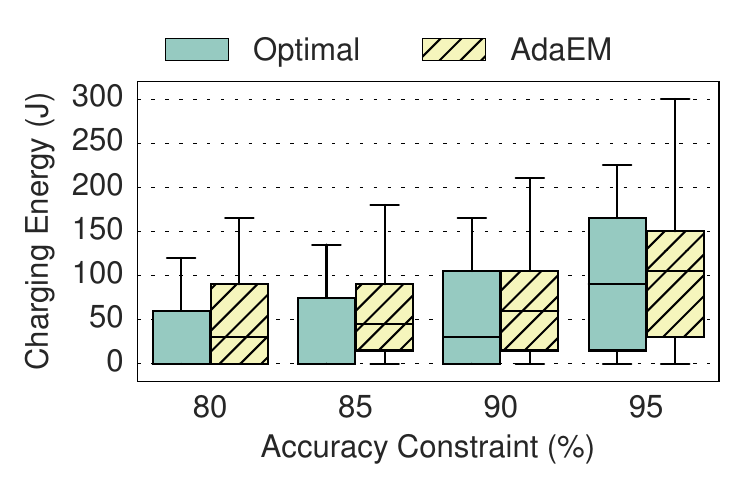}
	\caption{Distribution of charging energy as a function of the minimum activity constraint}
	\label{fig:charging_accuracy}
\end{figure}

The accuracy and activity constraint are important factors in determining the energy consumption and the charging schedule.
If the minimum accuracy constraint is higher, the application needs to consume higher energy to maintain the accuracy. This, in turn, leads to higher charging requirements on the device.
The distribution of charging energy with the minimum accuracy constraint is shown in Figure~\ref{fig:charging_accuracy}. We see that the charging requirement is typically less than 100~J when the accuracy constraint is 80\%. The requirement increases to about 150~J when the accuracy constraint is 95\%. 
Overall, the distribution of the charging energy for the proposed approach closely follows the optimal solution.

\end{document}